\documentstyle[12pt]{article}

\begin{document}

\def\one{\!\!{\hbox{ 1\kern-.8mm l}}}
\newcommand{\hepth}[1]{{\tt hep-th/#1}}

\renewcommand{\theequation}{\thesection .\arabic{equation}}
\renewcommand{\section}[1]{\setcounter{equation}{0}
 \addtocounter{section}{1}
 \vspace{5mm} \par \noindent {\large \bf \thesection . #1}
 \setcounter{subsection}{0} \par \vspace{2mm} } 
\newcommand{\sectionsub}[1]{\addtocounter{section}{1}
 \vspace{5mm} \par \noindent {\bf \thesection . #1}
 \setcounter{subsection}{0}\par}
\renewcommand{\subsection}[1]{\addtocounter{subsection}{1}
 \vspace{2.5mm}\par\noindent {\bf  \thesubsection . #1}\par
 \vspace{0.5mm}}
\renewcommand{\thebibliography}[1]{{\vspace{5mm}\par 
 \noindent{\large \bf References}\par \vspace{2mm}} 
 \list{$[$\arabic{enumi}$]$}{\settowidth\labelwidth{[#1]}
 \leftmargin \labelwidth \advance\leftmargin\labelsep
 \addtolength{\topsep}{-4em}\usecounter{enumi}}
 \def\newblock{\hskip .11em plus .33em minus .07em}
 \sloppy\clubpenalty4000\widowpenalty4000 \sfcode`\.=1000\relax 
 \setlength{\itemsep}{-0.4em}}
\newcommand{\acknowledgments}[1]{\vspace{5mm}\par 
 \noindent{\large \bf Acknowledgments}\par \vspace{2mm}}

\def\a{& \hspace{-7pt}}
\def\c{\hspace{-5pt}}
\def\bea{\begin{eqnarray}}
\def\eea{\end{eqnarray}}
\def\be{\begin{equation}}
\def\ee{\end{equation}}
\def\nn{\nonumber}
\def\alp{\alpha}
\def\bet{\beta}
\def\gam{\gamma}
\def\del{\delta}
\def\eps{\epsilon}
\def\sig{\sigma}
\def\lam{\lambda}
\def\Lam{\Lambda}
\def\m{\mu}
\def\n{\nu}
\def\r{\rho}
\def\s{\sigma}
\def\d{\delta}
\def\Z{{\bf Z}}
\def\e{\epsilon}
\def\st{\scriptstyle}
\def\mco{\multicolumn}
\def\epp{\epsilon^{\prime}}
\def\vep{\varepsilon}
\def\ra{\rightarrow}
\def\ab{\bar{\alpha}}

\newcommand{\sect}[1]{\setcounter{equation}{0} \section{#1}}
\newcommand{\eqn}[1]{(\ref{#1})}
\newcommand\rf[1]{(\ref{#1})}
\newcommand{\NPB}[3]{{Nucl.\ Phys.} {\bf B#1} (#2) #3}
\newcommand{\CMP}[3]{{Commun.\ Math.\ Phys.} {\bf #1} (#2) #3}
\newcommand{\PRD}[3]{{Phys.\ Rev.} {\bf D#1} (#2) #3}
\newcommand{\PLB}[3]{{Phys.\ Lett.} {\bf B#1} (#2) #3}
\newcommand{\JHEP}[3]{{JHEP} {\bf #1} (#2) #3}
\newcommand{\ft}[2]{{\textstyle\frac{#1}{#2}}\,}
\newcommand{\dt}{\partial_{\langle T\rangle}}
\newcommand{\dtbar}{\partial_{\langle\bar{T}\rangle}}
\newcommand{\al}{\alpha^{\prime}}
\newcommand{\mst}{M_{\scriptscriptstyle \!S}}
\newcommand{\mpl}{M_{\scriptscriptstyle \!P}}
\newcommand{\dv}{\int{\rm d}^4x\sqrt{g}}
\newcommand{\lv}{\left\langle}
\newcommand{\rv}{\right\rangle}
\newcommand{\ph}{\varphi}
\newcommand{\sbar}{\,\bar{\! S}}
\newcommand{\xbar}{\,\bar{\! X}}
\newcommand{\fbar}{\,\bar{\! F}}
\newcommand{\zbar}{\,\bar{\! Z}}
\newcommand{\tbar}{\bar{T}}
\newcommand{\ybar}{\bar{Y}}
\newcommand{\phb}{\bar{\varphi}}
\newcommand{\cm}{Commun.\ Math.\ Phys.~}
\newcommand{\pr}{Phys.\ Rev.\ D~}
\newcommand{\prl}{Phys.\ Rev.\ Lett.~}
\newcommand{\pl}{Phys.\ Lett.\ B~}
\newcommand{\ibar}{\bar{\imath}}
\newcommand{\jbar}{\bar{\jmath}}
\newcommand{\np}{Nucl.\ Phys.\ B~}
\newcommand{\eqalign}[1]{\hspace{-10pt}\begin{array}{ll} #1 
 \end{array}\hspace{-10pt}}
\newcommand{\dfrac}[2]{\displaystyle{\frac {#1}{#2}}}
\newcommand{\gsi}{\,\raisebox{-0.13cm}{$\stackrel{\textstyle>}
 {\textstyle\sim}$}\,}
\newcommand{\lsi}{\,\raisebox{-0.13cm}{$\stackrel{\textstyle<}
 {\textstyle\sim}$}\,}
\newcommand{\ds}[1]{\displaystyle{#1}}


\begin{titlepage}
\setcounter{page}{0}
\vskip 40pt
\begin{flushright}
LMU-TPW 99-22 \\
NEIP-99-018 \\
{\tt hep-th/9912090}
\end{flushright}

\vskip 80pt

\begin{center}
{\Large \bf On the effective action of stable non-BPS branes}

\vskip 20pt

{\large Rodolfo Russo$^{~a}$ and Claudio A. Scrucca$^{~b}$}

\vskip 20pt

${}^a$
{\em Institut de Physique, Universit\'e de Neuch\^atel}\\
{\em Breguet 1, CH-2000 Neuch\^atel, Switzerland}\\
{\footnotesize \tt Rodolfo.Russo@iph.unine.ch}

\vskip 5pt

${}^b$
{\em Sektion Physik, Ludwig Maximilian Universit\"at M\"unchen}\\
{\em Theresienstra\ss e 37, D-80333 Munich, Germany}\\
{\footnotesize \tt Claudio.Scrucca@physik.uni-muenchen.de}

\vskip 30pt

\end{center}

\begin{abstract}

\vskip 20pt

We study the world-volume effective action of stable non-BPS branes
present in Type II theories compactified on K3. 
In particular, by exploiting the conformal description of these
objects available in the orbifold limit, we argue that their 
world-volume effective theory can be chiral. 
The resulting anomalies are cancelled through the usual 
inflow mechanism provided there are anomalous couplings, 
similar to those of BPS branes, to the twisted R-R fields.
We also show that this result is in agreement with the conjectured 
interpretation of these non-BPS configurations as BPS branes wrapped 
on non-supersymmetric cycles of the K3. 

\end{abstract}

\end{titlepage}

\newpage


\section{Introduction}

In a remarkable series of papers \cite{Sen1,Sen2,Sen3,Sen6},
A. Sen showed that Type II string theories contain, besides D-branes, 
other extended objects which are not supersymmetric. These non-BPS 
branes can be viewed as the bound state formed by two coincident 
D-branes carrying opposite Ramond-Ramond charge when the world-volume
tachyon is condensed in a real kink solution. In this way, starting 
from two BPS D(p+1)-branes, one can describe a single 
non-BPS $p$-brane. Alternatively \cite{Sen6}, it is possible to describe 
the same non-BPS configuration by modding out the theory by the operator 
$(-1)^{F_L}$, whose effect is to change the sign of all R-R and R-NS states. 
In this case, the system of a D$p$-brane and a $\bar{\rm D}p$-brane in the
Type IIA (IIB) theory becomes a non-BPS $p$-brane in the Type IIB (IIA) 
theory\footnote{For the details of the different constructions we refer to 
the reviews~\cite{Senr,leru,schw1}.}. 
{}From this second point of view, a non-BPS brane can be defined as a 
hyper-plane where {\em two} different kinds of open strings, distinguished
by a Chan-Paton factor, can end. The first sector, with C-P factor $\one$,
is identical to the one living on usual D-branes, whereas the second 
sector, with C-P sector $\sigma_1$, differs from the first one in 
that it contains only states which are {\em odd} under the GSO operator 
$(-1)^F$. Due to this non-standard projection, a real open string tachyon
survives in the $\sigma_1$ sector, and a non-BPS brane is therefore unstable 
in a flat Type II theory. 

Interestingly, it is possible to construct stable states from the above
configurations by exploiting discrete symmetries of the
original Type II theory, under which the tachyonic field is odd. 
So far, two cases have been thoroughly studied: Type I theory, that is 
essentially Type IIB theory modded out by the world-sheet parity operator
$\Omega$ \cite{Sen4,Sen5,Gallot}; second, Type II theories compactified 
on the orbifold $T^4/\Z_2$ \cite{Sen3,Sen6,Gab1,Gab2}, the $\Z_2$ being 
generated by the reflection ${\cal I}_4$ of the four compact directions or 
its T-dual version ${\cal I}_4 (-1)^{F_L}$. These stable objects are 
interesting for several reasons. First, they are part of the spectra of 
the above string theories, and a description which misses them would be 
incomplete. Moreover, despite the fact that these states do not preserve 
any supersymmetry, they are simple enough to allow for a precise analysis
of their physical properties, like masses or couplings.
Indeed, non BPS-branes are much on the same footing as usual 
D-branes: their exact microscopic description is given by the conformal 
field theory of the open strings ending on them. 
{}From the closed string point of view, the properties of this 
conformal field theory can be resumed in the boundary state 
formalism~\cite{clny88}, in which D-branes are described by a coherent 
closed string state inserting a boundary on the string world-sheet and 
enforcing on it the appropriate boundary conditions. The boundary state 
approach can be naturally extended also to non-BPS 
branes~\cite{Gab1,Sen2,Gallot}, and is often a very useful tool for 
describing branes from the point of view of bulk theory. 

The effective action for non-BPS brane has
been studied by Sen in Ref.~\cite{Sen7} mainly in flat type II
context or, in the stable case, by restricting to couplings to states of
the untwisted closed string sector.

In this letter, we study additional couplings to states in the twisted sectors
arising at the fixed-points where the curvature is concentrated. 
For convenience, we focus on the standard ${\cal I}_4$ orbifold\footnote{For 
a detailed analysis of the perturbative and the D-brane spectrum in this 
theory see \cite{gabstef}.}, but in the $T$-dual case 
${\cal I}_4 (-1)^{F_L}$ one recovers similar results.
In particular, we show that the non-BPS brane action includes a 
Wess-Zumino term involving twisted R-R states. Beside the expected minimal 
coupling to the appropriate form, this contains also anomalous couplings 
to lower forms. We will show that these couplings induce a tree-level inflow 
which compensates one-loop anomalies that can arise in the world-volume 
theory; they are therefore crucial for the consistency of the theory. 
We also show that the appearance of such anomalous couplings for these 
stable non-BPS branes is in agreement with their interpretation as BPS 
branes wrapped on non-supersymmetric cycles.


\section{Non-BPS branes in Type II on $T^4/\Z_2$}

In Ref.~\cite{Sen6} it was pointed out that a non-BPS brane with an
odd number of directions wrapped on the orbifold $T^4/{\cal I}_4$ is a
stable object in a certain region of the moduli space. Let us briefly
recall under what conditions the tachyonic field disappears. 
Consider a non-BPS $(p+n)$-brane with $n$ (odd) Neumann directions
in the compact space ($x^6,x^7,x^8,x^9$). As usual in toroidal 
compactification, open string states living on the brane can have 
Kaluza-Klein modes along Neumann directions ($x^a$) and winding modes 
along Dirichlet ones ($x^i$). The effective mass of these states is
\be \label{mass}
m^2 = \sum_a \left(\frac{n_a}{R_a}\right)^2 + \sum_i \left({w^i R_i\over
\alpha'} \right)^2 - {1\over 2\alpha'} \;.
\ee
If the radii of the compact dimensions satisfy the relations 
$R_a \leq \sqrt{2\alpha'}$ and $R_i \geq \sqrt{\alpha'/2}$, 
the only open string state which is really tachyonic is the
zero-mode $n_a=w^i=0$. Clearly, this instability can not be cured by 
adjusting the value of the moduli in a simple toroidal compactification,
and in order to stabilize the non-BPS branes the $\Z_2$-projection
plays a crucial role. 

Let us first recall some generalities of the $D=6$ $N=2$ bulk theory. 
In the untwisted closed string sector, modding out ${\cal I}_4$ kills half 
of the original physical degrees of freedom. 
One is then left with a gravitational multiplet and either 4 
vector multiplets of $N=(1,1)$ supersymmetry for the Type IIA theory and 
5 tensor multiplet of $N=(2,0)$ supersymmetry for the Type IIB theory.
At each of the 16 orbifold fixed-planes, there are also twisted sectors, 
in which strings close up to an ${\cal I}_4$ identification. It turns
out that, in this case, one recovers a supersymmetric spectrum by
using, also in the twisted sector, the natural GSO projection and by
keeping the even states under ${\cal I}_4$ \cite{gabstef}. 
At each of the orbifold fixed-points, one gets a vector multiplet in the 
Type IIA case and a tensor multiplet in the Type IIB case.

In order to discuss the low-energy world-volume theory on the previously 
introduced non-BPS brane, one has then to define the $\Z_2$ action on the 
open string states living on it, and in particular on their C-P 
wave-function. The correct procedure is to impose the conservation of the 
quantum numbers with respect to ${\cal I}_4$ in the interactions among open 
and closed strings. In the $\one$ sector, the natural choice of taking 
the open string vacuum as an even state under ${\cal I}_4$ turns out to be 
consistent. Conversely, in the $\sigma_1$ sector, one is forced to take the 
opposite choice. To see this, it is sufficient to look at the two-point 
amplitude between an untwisted R-R state and an open string tachyon. 
This is the well-known coupling \cite{WITTEN,Sen6,HORAVAK,bcr} of a
non-BPS brane to the untwisted forms arising when the tachyon is not
constant $\int dT\wedge C^{(p)}$. In momentum space this gives a
vertex of the form $k_T^0 \, C^{\mu_1\ldots\mu_p\mu_{p+1}
\ldots\mu_{p+n}}$; since $C$ (like the brane emitting it) has an odd
number of Lorentz indices in the compact space, it is 
odd under ${\cal I}_4$ and thus also the tachyon field has to be so.
This means that the tachyonic zero mode is projected out and the
non-BPS brane is stable when one considers the range of radii
specified after Eq.~(\ref{mass}). Thus, one can resume the projection
rules on the open string sectors by saying that the $\Z_2$ operation
acts also on the C-P factor adding a minus sign to the states
in the $\sigma_1$ sector. 

{}From the closed string point of view these non-BPS brane are described
by a boundary state containing only the untwisted part of the NS-NS
sector and the twisted part of the R-R sector \cite{Gab1,Sen4}. Since
the boundary state encodes the couplings of the brane with all the
states of the closed string spectrum (see e.g. \cite{bs}), we can conclude 
that the effective action has to contain a DBI part, describing the couplings
to NS-NS untwisted states, and a WZ part, encoding the interactions
with twisted R-R states. As usual, the orbifold projection does not change 
the couplings among the fields in the untwisted sector, thus one can 
read this part of the action from the result obtained in flat Type II  
theory \cite{Sen7} by simply setting to zero the fields which are odd 
under ${\cal I}_4$. On the contrary the WZ part, involving twisted fields,
has to be explicitly calculated and, as we shall see in the next
section, the result contains couplings to lower (twisted) forms
which are related to an inflow of anomaly. This may seem strange,
since non-BPS branes contain twice as many fermions as the usual
D-branes and usually in type II theories the two sets come with opposite
chirality. However, by analyzing carefully the effect of the $\Z_2$
projection on the Ramond sector, one sees that a chiral open string 
spectrum may emerge. 

Consider first the spacetime filling case $p=5$. In this case, the 
ten-dimensional Lorentz group $SO(9,1)$ is broken to $SO(5,1) \times SO(4)$.
In the $\one$ sector, the standard GSO-projection leads to an $N=2$ gauge 
multiplet. In the bosonic sector, this contains one gauge boson in the 
$({\bf 6}, {\bf 1})$ and 4 scalar fields in the $({\bf 1}, {\bf 4})$, 
corresponding to the dimensional 
reduction of a gauge boson from $D=10$. In the fermionic sector, 
there are spinors in the $({\bf 4}, {\bf 2}) \oplus 
({\bf 4'}, {\bf 2'})$, corresponding to the dimensional reduction 
of a chiral Majorana-Weyl spinor ${\bf 16}$ from $D=10$. 
In the $\sigma_1$ sector, the non-standard GSO-projection leads instead 
to a bosonic tachyon and spinors in the 
$({\bf 4}, {\bf 2'}) \oplus ({\bf 4'}, {\bf 2})$, corresponding to 
the dimensional reduction of an anti-chiral Majorana-Weyl spinor 
${\bf 16'}$ from $D=10$. Finally, one has to keep only $\Z_2$-invariant
states; but since the projection also depends on the C-P
factor, this means that one has to keep ${\cal I}^{0}_4$-even states
in the $\one$ sector and ${\cal I}^{0}_4$-odd states in the $\sigma_1$
sector, where ${\cal I}^{0}_4$ is the orbital contribution to the $\Z_2$
orbifold operator (without the C-P contribution). As already
discussed, the tachyon is odd under the global ${\cal I}_4$, and is 
therefore projected out. Moreover, the ${\bf 4}$ of $SO(4)$ is by 
construction odd under ${\cal I}^0_4$, and using the decomposition 
${\bf 4} = {\bf 2} \otimes {\bf 2'}$, one concludes that if the 
${\bf 2}$ is chosen to be even under ${\cal I}^0_4$, then the ${\bf 2'}$
has to be odd. The surviving spectrum is therefore found to consist of a 
gauge boson $({\bf 6}, {\bf 1})$ ($\one$ sector) and the spinors 
$({\bf 4}, {\bf 2})$ ($\one$ sector) and $({\bf 4}, {\bf 2'})$
($\sigma_1$ sector). 
In order to check that this projection is correct
also in the R open string sector, it is sufficient to consider
the 3-point amplitude between a tachyon and two massless fermions, the
first in the $\one$ sector and the second in the $\sigma_1$ sector.
This amplitude is proportional to ${\rm Tr}(\sigma_1^2)\,\langle 0|\;
S_{-1/2}^{\dot\alpha} \,{\rm e}_{-1}^{{\rm i} k \cdot X}\,S_{-1/2}^{\beta} 
|0\rangle$, and does not vanish when the first spinor is in the
$({\bf 4}, {\bf 2})$ and the second is in the $({\bf 4'}, {\bf 2})$,
because the product of these representations contains the
ten-dimensional scalar. This shows that the two spinors have opposite 
eigenvalue under ${\cal I}_4$ and that one can not keep both of them in 
the orbifolded theory. The non-supersymmetric world-volume theory is 
therefore chiral.

In the cases $p<5$, the six-dimensional Lorentz group $SO(5,1)$ is further 
broken to $SO(p,1) \times SO(5-p)$. The world-volume theory is 
obtained by simple dimensional reduction of the world-volume theory for 
the $p=5$ case from 6 to $p+1$ dimensions.


\section{Anomalies}

We have seen in previous section that the resulting world-volume 
theory is potentially chiral.
More precisely, the theory is strictly chiral only for the spacetime 
filling case $p=5$ (non-BPS 6,8-brane wrapped on 1,3 directions).
For the cases $p=3$ and $p=1$, the world-volume theory is obtained 
by dimensional reduction and is correspondingly 
non-chiral, since fermions decompose into two sets of
opposite chiralities. However, due to the fact that the original 
fermions were chiral, these two sets of spinors transform as fermionic
representations of opposite chiralities also with respect to the group 
of transverse rotations $SO(5-p)$, and a chiral asymmetry is generated 
when the normal bundle to the brane (in the non-compact spacetime) 
is non-trivial.

Therefore, as happens in the BPS case, also non-BPS branes 
can have an anomalous 
world-volume theory, despite the fact that their string theory 
construction is perfectly well-defined. This is nothing but a 
particular case of the well known fact that a topological defect 
in a consistent quantum field theory leads to an apparent 
(local) violation of charge conservation if it supports fermionic 
zero modes \cite{ch}. Obviously, if the starting theory is consistent, 
this can not be the final result, and actually it happens that the 
topological defect develops suitable anomalous couplings to bulk fields, 
leading to an inflow of charge from the bulk. In other words, the 
world-volume one-loop anomaly is exactly canceled by a tree-level 
anomaly, and charge conservation is restored \cite{ch}.

The occurrence of R-R anomalous couplings for BPS topological defects like
D-branes or O-planes in string theory vacua is by now well established. 
These couplings are completely determined through the requirement that the 
inflow of anomaly associated to the corresponding magnetic interactions cancel 
all possible world-volume anomalies \cite{ghm,cy,ss1} (see also \cite{ss3}). 
They have been also determined through direct string theory computations 
\cite{mss,cr1,cr2,stef}.
Since also non-BPS branes potentially support anomalies, it is natural to 
expect that they will also develop anomalous couplings, and indeed we will
show that they do so. We shall follow the approach of \cite{ss2}, and 
extract the R-R anomalous by factorization from a string theory 
computation of the anomaly and the inflow.

The one-loop partition function on the non-BPS D-brane is given by the 
projected annulus vacuum amplitude
\be
Z(t) = \frac 14\,{\rm Tr}_{R-NS}\,[(1 + {\cal I}_4)(1 + (-1)^F)\,e^{-tH}] \;,
\ee
where $H$ is the open-string Hamiltonian and the trace contains
a sum over the two C-P sectors $\one$ and $\sigma_1$.
The generating functional of one-loop correlation functions of photons
and gravitons on the non-BPS D-brane is obtained by integrating with the 
correct measure over the modular parameter $t$ of the annulus the 
above partition function, evaluated in a gauge and gravitational background: 
$\Gamma = \int_0^\infty dt \,\mu(t)\, Z(t)$. 
Clearly, possible anomalies can emerge only in the CP-odd part of this 
effective action, associated to the odd spin-structure, and happen to be 
boundary terms in moduli space.

In \cite{ss2}, a general method to compute directly the anomalous part 
of the effective action through an explicit string computation has 
been presented. The gauge variation is represented by the insertion 
of an unphysical vertex in the amplitude, representing a photon
or a graviton with pure-gauge polarization. After formal 
manipulations, this unphysical vertex combines with the world-sheet 
supercurrent appearing in odd spin-structure amplitudes, to leave
the $t$-derivative of the correlation of a certain effective 
vertex operator in a generic gauge and gravitational background. 
Interestingly, the effect of this operator was recognized 
to correspond to obtaining the anomaly as Wess-Zumino descent, 
$A = 2 \pi i \int I^{(1)}$, the anomaly polynomial $I$ being given 
by the background-twisted partition function. We use here the 
standard descent notation: given a gauge-invariant polynomial $I$ 
of the gauge and gravitational curvatures $F$ and $R$, one defines
$I^{(0)}$ such that $I = d I^{(0)}$ and $I^{(1)}$ through the gauge 
variation $\delta I^{(0)} = d I^{(1)}$.

In consistent string vacua, only the UV boundary $t \rightarrow 0$ 
can potentially lead to anomalies, and it turns out that this vanishes by
itself.
At low-energy, this is interpreted as Green-Schwarz mechanism, through which 
the quantum one-loop anomaly is cancelled by an equal and opposite 
classical tree-level inflow. It was suggested in \cite{ss2} that such an
interpretation can be recovered by taking the limit of slowly varying
background fields, corresponding to low momenta for external particles in the 
anomalous graph. In this limit, the partition function becomes a topological 
index which is independent of the modulus $t$, and one ends up with the 
extremely simple recipe that the polynomial $I$ from which both the anomaly 
and the inflow descend (a ($D$+2)-form in $D$ dimensions) is given simply by 
this partition function, with the bosonic zero modes excluded and the 
convention of working in two dimensions higher.

In our case, the anomaly polynomial is given by 
\be
I = \frac 14\,{\rm Tr}_R\,[(1 + {\cal I}_4)(-1)^F\,e^{-tH(F,R)}] \;,
\ee
where from now on it will be understood that one has to keep 
only the ($D$+2)-form component, with $D=p$ in our case. 
Recall now that both ${\cal I}_4$ and $(-1)^F$ act with opposite signs
in the two C-P sectors $\one$ and $\sigma_1$. The $(-1)^F$ 
part gives a vanishing contribution, due both to a cancellation 
between the two C-P sectors and to the four fermionic zero 
modes in the compact directions, which are not twisted by the background. 
The ${\cal I}_4 (-1)^F$ gives instead a non-vanishing result, since the 
above fermionic zero modes are absent and the two C-P sectors 
give exactly the same result, generating a factor of 2. 
For the rest, the computation of the partition function proceeds 
exactly as in \cite{mss,ss1,ss2}. The internal part 
contributes\footnote{These numbers can be obtained using $\zeta$-function 
regularization, as in \cite{ss2}.} a factor of $4$: the $n$ Neumann bosons 
give a factor $2^{-n}$ which cancels the fixed points degeneracy $2^n$, 
whereas each of the four fermions contributes $\sqrt{2}$. 
Setting $4 \pi^2 \alpha^\prime = 1$, one finds finally
\be
I(F,R,R^\prime) = 2 \, {\rm ch} (F) \, {\rm ch} (-F) \wedge 
\frac {\widehat{A}(R)}{\widehat{A}(R^\prime)}
\wedge e(R^\prime) \;. \label{ano}
\ee
Here $F$, $R$ and $R^\prime$ indicate the curvature forms of the gauge, 
tangent and normal bundles. $\widehat{A}(R)$ and $\widehat{A}(R^\prime)$ 
are the Roof genera of the tangent and the normal bundles, $e(R^\prime)$ 
is the Euler class of the normal bundle (defined to be $1$ when the latter 
is null), and ${\rm ch} (F)$ is the Chern character of the gauge bundle in
the fundamental representation.

This result gives both the one-loop anomaly on the non-BPS D-brane
and the opposite inflow that cancels it. The latter is related 
to the presence of anomalous couplings of the non-BPS D(p+n)-brane to
the R-R fields in the twisted sectors arising at the $2^n$ fixed-points
contained in the $n$ compact directions of the world-volume. 
The general from of these couplings is:
\be
S = - \raisebox{2pt}{$\displaystyle{\sum_i}$}\,
\frac {\mu_i}2 \int C_i \wedge Y_i(F,R,R^\prime) \;,
\ee
where $i=1,...,2^n$ labels the fixed-points. $\mu_i$ is an arbitrary 
charge and $Y_i(F,R,R^\prime)$ a polynomial of the curvatures \footnote{We 
use the conventions of \cite{cy}, and normalize $\mu_i$ such that 
$Y_i(0) = 1$.}. Finally, $C_i$ is the formal sum of all the twisted R-R 
forms and their duals in the $i$-th sector. More precisely, in the relevant 
Type IIB case, one gets one $N=2$ tensor multiplet in each sector, with a 
R-R content of one scalar and one an anti-self-dual 2-form\footnote{This can 
be understood in the $N=1$ language of \cite{ss2}, where the $N=2$ tensor 
multiplet decomposes into a hypermultiplet containing 1 R-R and 3 NS-NS 
scalars, plus a tensor multiplet, containing a R-R anti-self-dual 2-form 
and a NS-NS scalar.}, and each $C_i$ is therefore the sum of forms of 
degree 0, 2 and 4.
In \cite{cy}, it was shown that the inflow generated by such anomalous 
couplings is given in modulus by
\be
I(F,R,R^\prime) = \raisebox{2pt}{$\displaystyle{\sum_i}$} \,
\frac {\mu_i^2}{4\pi} \, Y_i (F,R,R^\prime) \wedge Y_i (-F,R,R^\prime)
\wedge e(R^\prime) \;.
\label{inf}
\ee
Comparing Eq.~(\ref{ano}) with Eq.~(\ref{inf}), one finally 
extracts
\be
\mu_i = \pm \sqrt{\frac{8\pi}{2^n}} \;,\;\;
Y_i (F,R,R^\prime) = {\rm ch} (F) \wedge 
\sqrt{\frac {\widehat{A}(R)}{\widehat{A}(R^\prime)}}
\;.
\ee
As expected \cite{Sen3,Sen6}, in the $n=1$ case the charge $\mu_i$ with 
respect to each of the two $C_i$ is identical to the twisted charge of 
a fractional brane at the same orbifold point. The factorization leaves 
obviously an unimportant sign ambiguity in each $\mu_i$, that we will 
ignore.

We conclude therefore that the stable non-BPS object obtained by 
wrapping a non-BPS (p+n)-brane of Type IIB along $n$-directions of 
$T^4/\Z_2$ has the anomalous couplings
\be
S = - \frac \mu 2 \int C \wedge {\rm ch} (F) \wedge 
\sqrt{\frac {\widehat{A}(R)}{\widehat{A}(R^\prime)}}
\label{cou}
\ee
where
\be
\mu = \sqrt{8\pi} \;,\;\; C = \frac 1{\sqrt{2^n}} \sum_{i=1}^{2^n} C_i \;,
\ee

Notice that, strictly speaking, the gauge bundle is restricted to have 
structure group $U(1)$ in the general construction above. To construct 
configurations with a $U(N)$ bundle, one would have to take $N$ wrapped 
non-BPS branes on top of each other, but unfortunately such a configuration 
is unstable since non-BPS branes repulse each other. However, it was 
pointed out in \cite{SenGab} that when all the radii take the critical
value, such force vanishes at the leading one-loop 
level, due to an accidental boson-fermion degeneracy in the world-volume 
theory. In this particular case, the configuration of $N$ overlapping 
non-BPS (p+n)-branes becomes stable, and has indeed the anomalous coupling 
(\ref{cou}).


\section{Discussion}

It is well known that the ${\cal I}_4$ orbifold describes a
particular limit of a smooth K3 manifold where 16 of the 22 2-cycles
are shrinking in cone singularities which correspond to the twisted
sectors arising in the conformal analysis. The result of previous
section has a natural interpretation also from this geometrical point
of view. In fact, in Ref.~\cite{Sen6} Sen proposed that the non-BPS
branes considered here can be viewed as BPS D-branes wrapped on
particular non-supersymmetric cycles present in the K3. This is very 
similar to the interpretation of fractional D-branes at fixed-points 
given in \cite{doumoore}: they can be viewed as ten-dimensional D-branes 
wrapped on the exceptional cycles coming from the resolution of orbifold 
singularities. More precisely, it is known that non-BPS branes decay, 
outside the stability region in the moduli space, into a D-brane and 
an anti-D-brane (i.e. with opposite R-R untwisted charge) wrapped on 
two supersymmetric cycles. This implies that the two configurations, 
the non-BPS brane and the D-brane - anti-D-brane pair into which it 
can decay, have to carry the same R-R charge.

{}From a geometrical point of view, where one interprets these
six-dimensional branes as ten-dimensional objects wrapped on cycles 
in K3, this statement means that the cycle giving rise to the non-BPS 
brane and the ones producing the BPS branes are in the same homology class. 
In the $n=1$ case, the latter are simply the two exceptional 2-cycles, 
coming from the resolution of the singularities touched by the brane. 
Interestingly, this is enough to deduce the couplings of the non-BPS brane 
from those of fractional branes. Indeed, the WZ action of the non-BPS 
has to come from the usual WZ action of the ten-dimensional BPS-brane, 
reduced on the appropriated 2-cycle. Fortunately, the integration over 
the cycle is in this case easy: the WZ part of the action contains 
only forms and so the only relevant thing of the considered cycle is 
its homology class. Using then the above homological decomposition, one 
finds that the WZ action for the non-BPS brane is given by the sum of 
the WZ action of a BPS brane and that of an anti-brane, wrapped on 
the two exceptional cycles. These are fractional branes, and the 
integration over the compact space must then give for each of these the 
action studied in \cite{ss1,ss2}\footnote{It should be possible to check
through an explicit calculation that the twisted sector couplings arise 
by decomposing the R-R forms on the exceptional 2-cycle, whereas the 
untwisted sector couplings comes from integrating the NS-NS two-form, which 
is known to have a non-vanishing flux across the cycle \cite{asp,nw}.}. 
The $n=3$ case is related to the case just discussed by T-duality, 
and thus the general pattern is the same. The only difference is that
the BPS brane configuration in which the non-BPS brane can decay present
more complicated supersymmetric 2-cycles. 
In both cases, summing all the WZ actions for the BPS branes, the untwisted 
part cancels, and for the twisted part one gets precisely the result of 
Eq.~(\ref{cou}). This provides strong evidence that both the geometrical 
interpretation of \cite{Sen6} and the anomalous couplings derived here 
are correct.

As was already the case for BPS D-branes, the part of the action
involving R-R field presents some particular features: from the open
string point of view the result is determined by an odd spin-structure 
amplitude, where only an effective zero mode part of the various vertices 
is relevant for the final result. This is related to the fact that 
from the field theory point of view, these couplings are related to
possible 1-loop anomalies in the world-volume theory. Finally from the
geometrical point of view, the WZ term is determined simply by the
homology class of the 2-cycle defining the non-BPS brane.

The DBI part of the action does not share these simplifying features
and so can not be determined directly with the above techniques. The
only simple observation we can make at this point is that there is no
linear coupling to twisted fields. In absence of a gauge field on the 
brane, this is evident from the form of the boundary state
written in \cite{Gab1,Sen6} which does not present the NS-NS twisted
part. The same conclusion, of course, can be derived from an annulus
calculation along the lines of \cite{abd}, by focusing on the 
NS$\;{\cal I}_4$ sector. As pointed out in Ref.~\cite{SenGab}, this 
part gives a vanishing contribution due to a cancellation between the 
two open string sectors ($\one$ and $\sigma_1$). 
This calculation is not essentially modified by the
introduction of a gauge field on the non-BPS brane: in fact, both
sectors are charged under the relative $U(1)$ field and the cancellation
between the two sectors still holds, showing that no linear coupling to
the NS-NS twisted scalars are produced by turning on a gauge 
field\footnote{We are grateful to C. Bachas for important discussions
on this point.}.
A possible way to go beyond this approximation is to use the geometrical 
interpretation of non-BPS branes and to derive not only the homology 
class of the 2-cycle defining them, but also its shape. Work is in 
progress in this direction.


\vspace{5mm}
\par \noindent {\large \bf Acknowledgments}
\vspace{3mm}

We would like to thank A.~Sen for useful comments and for reading this
manuscript, and C. Bachas, A. Bilal, M. Bill\'o, C.-S. Chu, M. Frau, 
A. Lerda, E. Scheidegger, M. Serone for interesting discussions 
and suggestions. C.A.S. is also grateful to the Physics Institute of the 
University of Neuch\^atel for hospitality. This work has been supported 
by the EEC under TMR contract ERBFMRX-CT96-0045 and the Fond National Suisse.


\end{document}